\documentclass[12pt]{article}
\usepackage[utf8]{inputenc}
\usepackage{slashed}
\usepackage{epsfig}
\usepackage{comment}
\usepackage{feynmp}
\usepackage{cancel}
\usepackage{bbm}

\usepackage{amsfonts}
\usepackage{amsmath}
\usepackage{amssymb}
\usepackage{array}
\usepackage{bigints}
\usepackage{booktabs}
\usepackage[nosort]{cite}
\usepackage{color}
\usepackage{dsfont}
\usepackage{float}
\usepackage{framed}
\usepackage{graphicx}
\usepackage{indentfirst}
\usepackage{mathrsfs}
\usepackage{multirow}
\usepackage{setspace}
\usepackage{subdepth}
\usepackage{subfig}
\usepackage{titlesec}
\usepackage[dotinlabels]{titletoc}
\usepackage{wrapfig}
\usepackage[all]{xy}
\usepackage{young}
\usepackage[vcentermath]{youngtab}
\usepackage{relsize}
\usepackage{hyperref}
\hypersetup{colorlinks=true}
\hypersetup{linkcolor=black}
\hypersetup{citecolor=black}
\hypersetup{urlcolor=black}
\numberwithin{equation}{section}

\def\e{{\epsilon}}

\def\p{\partial}

\def\0{{(0)}}
\def\1{{(1)}}
\def\2{{(2)}}

\def\<{\langle }
\def\>{\rangle }

\def\p{\partial}

\newcommand{\bea}{\begin{eqnarray}}
\newcommand{\eea}{\end{eqnarray}}
\newcommand{\ba}{\begin{align}}
\newcommand{\ea}{\end{align}}

\newcommand{\bigO}{\mathcal{O}}

\newcommand{\tr}[1]{\text{tr} \left[ #1 \right]}

\newcommand{\beq}{\begin{equation}}
\newcommand{\eeq}{\end{equation}}
\newcommand{\beqa}{\begin{eqnarray}}
\newcommand{\eeqa}{\end{eqnarray}}
\newcommand{\beqar}{\begin{eqnarray*}}

\newcommand{\ie}{{ i.e.}\ }

\def\[{\[}
\def\]{\]}

\def\a{{\alpha}}

\def\la{\langle}
\def\ra{\rangle}

\newcommand{\bd}[1]{\begin{fmffile}{#1}\begin{fmfgraph*}}
\newcommand{\ed}{\end{fmfgraph*}\end{fmffile}}

\def\tilde{\widetilde}

\def\1{{\mathds 1}}

\def\ie{\begin{equation}\begin{aligned}}
\def\fe{\end{aligned}\end{equation}}

\usepackage[left=2.5cm,right=2.5cm,top=2.5cm,bottom=3cm]{geometry}
\makeatletter
\renewcommand\section{\@startsection {section}{1}{\z@}%
                                 {-3.5ex \@plus -1ex \@minus -.2ex}
                                   {2.3ex \@plus.2ex}%
                                   {\normalfont\large\bfseries}}
\renewcommand\subsection{\@startsection{subsection}{2}{\z@}%
                                   {-3.25ex\@plus -1ex \@minus -.2ex}%
                                     {1.5ex \@plus .2ex}%
                                     {\normalfont\bfseries}}
\renewcommand\subsubsection{\@startsection{subsubsection}{3}{\z@}%
                                   {-3.25ex\@plus -1ex \@minus -.2ex}%
                                     {1.5ex \@plus .2ex}%
                                     {\normalfont\itshape}}
\makeatother

\DeclareFontShape{OT1}{cmr}{mx}{n}%
    {<->cmr10}{}

\DeclareMathAlphabet{\titlemath}{OT1}{cmr}{mx}{n}

\begin{document}


\begin{titlepage}
\unitlength = 1mm
\ \\
\vskip 3cm
\begin{center}

{ \LARGE {\textsc{ Limitations on Dimensional Regularization in R\'enyi Entropy}}}

\vspace{0.8cm}

Ning Bao$^{1,2}$ and Temple He$^3$

\vspace{1cm}

\textit{$^1$Walter Burke Institute for Theoretical Physics, \\ California Institute of Technology, Pasadena, CA
91125, USA}
\\
\vspace{0.3cm}
\textit{$^2$Institute for Quantum Information and Matter, \\ California Institute of Technology, Pasadena, CA
91125, USA}
\\ 
\vspace{0.3cm}
\textit{$^3$Center for the Fundamental Laws of Nature, Department of Physics, \\
Harvard University, Cambridge, MA 02138, USA}\\

\begin{abstract}

Dimensional regularization is a common method used to regulate the UV divergence of field theoretic quantities. When it is used in the context of R\'enyi entropy, however, it is important to consider whether such a procedure eliminates the statistical interpretation thereof as a measure of entanglement of states living on a Hilbert space. We therefore examine the dimensionally regularized R\'enyi entropy of a 4d unitary CFT and show that it admits no underlying Hilbert space in the state-counting sense. This gives a concrete proof that dimensionally regularized R\'enyi entropy cannot always be obtained as a limit of the R\'enyi entropy of some finite-dimensional quantum system.

 \end{abstract}

\vspace{1.0cm}

\end{center}

\end{titlepage}

\newpage
\tableofcontents

\section{Introduction}

Quantum entanglement is a subject that is ubiquitious in all areas of physics, from condensed matter physics \cite{hastings2007area} to quantum information theory \cite{lieb2002proof} to, most recently, high-energy physics \cite{Ryu:2006bv, Bao:2015bfa}. There are many quantities that are utilized to measure the entanglement entropy between two systems $A$ and $B$. Arguably the most popular measurement is the von Neumann entropy:
\begin{align}
	S^A = \tr {\rho_A\log\rho_A},
\end{align}
where $\rho_A$ is the reduced density matrix obtained by partially tracing out system $B$. However, one can also choose instead to measure entanglement using the R\'enyi entropy \cite{Headrick:2010zt}:
\begin{align}\label{renyi}
 	S^A_n = \frac{1}{1-n}\log\tr{\rho_A^n}.
\end{align}
R\'enyi entropy is useful to study because in the limit $n \to 1$, we recover the von Neumann entropy, i.e. $S^A = \lim_{n\to 1} S^A_n$. Similarly, the holographic calculation of entanglement entropy is specifically an analaytic continuation of the replica trick as applied to $n$ copies of the R\'enyi entropy for $n=1$ \cite{Headrick:2010zt}. Finally, by knowing the R\'enyi entropy for all $n$, we can reconstruct the entire eigenvalue spectrum of $\rho_A$ \cite{Dong:2016wcf}.  We will henceforth drop the superscript $A$ and always implicitly assume our entangling region to be $A$.

In the context of high-energy physics, both the von Neumann and R\'enyi entropies diverge. In particular, given a $d$-dimensional conformal field theory (CFT), the R\'enyi entropy of a $(d-1)$-dimensional spatial region $A$ bounded by $\Sigma \equiv \p A$ is of the form\cite{Dong:2016wcf}
\begin{align}
	S_n = c_n^{(0)}\frac{\text{Area}(\Sigma)}{\e^{d-2}} + \cdots + S^{\text{univ}}_n + \tilde\a,
\end{align}
where $\e$ is the ultraviolet (UV) regulator. The set of dots corresponds to subleading power-law divergences, and $\tilde\a$ denotes the finite terms, including counterterms, that are theory-dependent. The term $S^{\text{univ}}_n$ is independent of the details of the UV cutoff and hence universal, whereas the other terms on the right-hand side are all scheme-dependent and hence nonuniversal. 

In order to tame these infinities, one method often employed in the literature is to simply use dimensional regularization and then drop the power law divergences using some subtraction scheme. The result is 
\begin{align}\label{renorm_renyi}
	S_n^{\text{DR}} = S^{\text{univ}}_n + \a,
\end{align}
where $\a$ may differ from $\tilde\a$ by finite counterterms introduced depending on the subtraction scheme. While this expression has been extremely useful in deepening our understanding of the nature of R\'enyi entropy and is widely utilized in the literature, it is also important to understand its limitations. Entanglement entropy and R\'enyi entropy are fundamentally quantities defined to measure the entanglement between quantum states. This is often called the state-counting interpretation of entropy, since we begin with a state living in a Hilbert space over regions $A$ and $B$, and by partially tracing out region $B$ we obtain the entanglement between $A$ and $B$. Clearly, if we choose to renormalize our theory by introducing a physical regulator like a UV cutoff, the state-counting interpretation is preserved. However, it is unclear whether there exists such an interpretation if we instead choose to use dimensional regularization, as was the procedure used to obtain \eqref{renorm_renyi}. This issue was explored in the context of entanglement entropy in \cite{Cooperman:2013iqr}. As we will demonstrate in this short paper, the dimensional regularization procedure does not offer such a state-counting interpretation in the case of a 4d unitary CFT, and there does not exist a Hilbert space on which we can define states such that we can derive \eqref{renorm_renyi} from \eqref{renyi}.

The structure of this paper is as follows. In section 2, we review the form of the R\'enyi entropy for a 4d CFT. In section 3, we will assume that the regulated R\'enyi entropy admits a state-counting interpretation, and hence in particular must satisfy certain entropy inequalities. This allows us to obtain a contradiction. In section 4, we summarize our results and discuss future potential directions.

\section{Universal R\'enyi Entropy Term in 4d}

We begin by constraining ourselves to a 4d spacetime, where $A$ is a 3d entangling spatial region, and $\Sigma = \p A$ its 2d boundary. Given any 4d CFT, the R\'enyi entropy $S_n^{\text{DR}}$ is fixed by Weyl invariance up to three functions of $n$ \cite{Bianchi:2015liz,Fursaev:2012mp,Solodukhin:2008dh}:
\begin{align}\label{myers}
 	S_n^{\text{DR}} = -\left(\frac{f_a(n)}{2\pi}\int_{\Sigma} R_\Sigma + \frac{f_b(n)}{2\pi}\int_{\Sigma} \tilde K_{ij}\tilde K^{ij} - \frac{f_c(n)}{2\pi}\int_{\Sigma}\gamma^{ij}\gamma^{kl}C_{ijkl} \right)\log(\mu l_A) + \a_\Sigma(n,l_A),
\end{align}
where $l_A$ is a characteristic length scale of $\Sigma$, $\mu$ an arbitrary mass scale used as an IR regulator often taken to be $\e^{-1}$, $\gamma^{ij}$ the inverse of the induced metric on the entangling surface, $R_\Sigma$ the Ricci curvature of the surface $\Sigma$, $C_{ijkl}$ the Weyl tensor, and $\tilde K_{ij}$ the traceless part of the second fundamental form given via
\begin{align}
 	\tilde K_{ij} \equiv K_{ij} - \frac{K}{2}\gamma_{ij},  \quad K \equiv K^{ij}\gamma_{ij}.
\end{align}
We have denoted the finite pieces collectively using $\a_\Sigma(n,l_A)$ to suggest that they potentially depend on the entangling surface $\Sigma$ as well as on $n$ and $l_A$. We will throughout this paper use Greek indices for tensors in 4d spacetime, and Latin indices for tensors induced on the 2d surface $\Sigma$.

In 4d, we have the relation 
\begin{align}\label{weyl_trace}
 	\gamma^{ij}\gamma^{kl}C_{ikjl} = \gamma^{ij}\gamma^{kl}R_{ikjl} - \gamma^{ij}R_{ij} + \frac{1}{3}R.
\end{align}
Hence, we see that we can express \eqref{myers} entirely in terms of the Riemann tensor of the background spacetime and the Riemann tensor and extrinsic curvature of the surface $\Sigma$. For the purposes of this paper, we will assume
\begin{align}\label{planck}
 	\log(\mu l_A) > 0,
\end{align}
which is equivalent to the assumption that the region $A$ under consideration is greater than Planck size. 

Let us define\footnote{We have in \eqref{lambda} suppressed the dependence $\lambda$ has on the spacetime metric on the left-hand side for notational simplicity. For the purposes of this paper we will only consider a Minkowski background.}
\begin{align}\label{lambda}
	\lambda(n,\Sigma) \equiv \frac{f_a(n)}{2\pi}\int_{\Sigma} R_\Sigma + \frac{f_b(n)}{2\pi}\int_{\Sigma} \tilde K_{ij}\tilde K^{ij} - \frac{f_c(n)}{2\pi}\int_{\Sigma}\gamma^{ij}\gamma^{kl}C_{ijkl}.
\end{align}
It follows we may write \eqref{myers} as
\begin{align}\label{myers_revised}
	S_n^{\text{DR}} = -\lambda(n,\Sigma)\log(\mu l_A) + \a_\Sigma(n,l_A).
\end{align}

\section{R\'enyi Entropy Constraints}

\subsection{R\'enyi Entropy Inequalities}

In this section, our goal is to demonstrate that $S_n^{\text{DR}}$ does not allow for a state-counting interpretation. Let us proceed via proof by contradiction and assume that $S_n^\text{DR}$ admits a state-counting interpretation. Then it in particular must satisfy the following inequalities for all $n>1$:
\begin{align}\label{ineq}
\begin{split}
	S_n^\text{DR} &\geq 0 \\
	 \frac{\p S_n^{\text{DR}}}{\p n} &\leq 0.
\end{split}
\end{align}
As we will show, the proofs for these are straightforward (see i.e. \cite{RevModPhys.50.221}). If $S_n^\text{DR}$ admits a state-counting interpretation, it can be written as
\begin{align}\label{renyi_2}
	S_n^\text{DR} = \frac{1}{1-n}\log\sum_{i=1}^N p_i^n,
\end{align}
where $p_i \in [0,1]$ for $i=1,\ldots,n$ and $\sum_{i=1}^N p_i = 1$. As $n > 1$, this implies $\sum_{i=1}^N p_i^N \leq 1$, with equality holding only if $p_{i_0} = 1$ for some $i_0$ and $p_i = 0$ for $i \not= i_0$. It follows immediately that $S_n^\text{DR} \geq 0$, proving the first inequality.

Similarly, the second inequality is proved by differentiating \eqref{renyi_2} with respect to $n$ to obtain
\begin{align}\label{renyi_deriv}
 	\frac{\p S_n^\text{DR}}{\p n} &= -\frac{1}{(1-n)^2}\sum_{i=1}^N P_i\log\frac{P_i}{p_i}, \quad 	P_i \equiv \frac{p_i^n}{\sum_{i=1}^N p_i^n}.
\end{align}
As the $P_i$'s sum to unity, the sum on the right-hand side of \eqref{renyi_deriv} can be thought of as the Kullback information gain, which is always positive. This completes the proof.

We now want to apply these two inequalities to the dimensionally regularized R\'enyi entropy \eqref{myers_revised}. Substituting \eqref{myers_revised} into \eqref{ineq}, we obtain
\begin{align}\label{constraint}
	 \lambda(n,\Sigma) &\leq  \frac{\a_\Sigma(n,l_A)}{\log(\mu l_A)} \\
	 \frac{\p \lambda(n,\Sigma)}{\p n} &\geq \frac{1}{\log(\mu l_A)}\frac{\p \a_\Sigma(n,l_A)}{\p n}.
\end{align}
Integrating the second line with respect to $n$, we get
\begin{align}\label{constraint_2}
	\lambda(n,\Sigma) - \lambda(n_0,\Sigma) &\geq \frac{\a_\Sigma(n,l_A) - \a_\Sigma(n_0,l_A)}{\log(\mu l_A)},
\end{align}		
where $n_0$ is a fixed integration constant.\footnote{We will always choose $n_0 > 1$ since we are only interested in R\'enyi entropy with $n > 1$.}

We now evaluate $\lambda(n,\Sigma)$ for a Minkowski background with $\Sigma$ being a sphere of radius $r$. The background metric is
\begin{align}\label{sch_metric}
 	ds^2  = -dt^2 + dr^2 + r^2d\theta^2 + r^2\sin^2\theta d\phi^2,
\end{align}
and the induced metric on $\Sigma$ is just the metric on a sphere:
\begin{align}
 	\gamma_{ij}dx^idx^j = r^2\,d\theta^2 + r^2\sin^2\theta\,d\phi^2.
\end{align}
In this case, the only nonzero term in \eqref{lambda} is the term involving $f_a(n)$. Noting $R_{\Sigma} = \frac{2}{r^2}$, we have
\begin{align}
	\lambda\left(n,S^2\right) = 4f_a(n).
\end{align}
Substituting into \eqref{constraint} and \eqref{constraint_2} and noting that $l_A = r$, we get
\begin{align}
	4f_a(n) &\leq \frac{\a_{S^2}(n,r)}{\log(\mu r)} \label{fa_1} \\
	4f_a(n) - 4f_a(n_0) &\geq \frac{\a_{S^2}(n,r) - \a_{S^2}(n_0,r)}{\log(\mu r)}. \label{fa_2}
\end{align}

\subsection{Constraint on $\a_{S^2}(n,r)$}

The purpose of this section is to first examine \eqref{fa_2} above and derive a constraint on $\a_{S^2}(n,r)$; this will later be useful in obtaining the required contradiction. In particular, we will prove that $\a_{S^2}(n,r)$ cannot increase arbitrarily as a function of $r$ if $S_n^\text{DR}$ admits a state-counting interpretation. 

Suppose the contrary, that $\a_{S^2}(n,r)$ grows without bound as a function of $r$. Because the R\'enyi entropy \eqref{myers_revised} must go to zero as subregion $A$ goes to zero, and the finite terms $\a_{S^2}(n,r)$ are independent of the cutoff scale $\mu$, they must go to zero separately as $r \to 0$.\footnote{The $\log(\mu r)$ term in \eqref{myers_revised} contains an IR regulator, which means care must be taken when taking the $r\to 0$ limit.} This means that each term in $\a_{S^2}(n,r)$ must be proportional to some positive power of $r$.

As we're assuming $S_n^\text{DR}$ admits a state-counting interpretation, \eqref{fa_2} in particular must be satisfied. Fixing an arbitrary $n_0 > 1$, consider first the case when the left-hand side of \eqref{fa_2} is negative for some $n$. Then it is obvious from the inequality that we must also have
\begin{align}
	\a_{S^2}(n,r) < \a_{S^2}(n_0,r), \quad\text{given} \quad f_a(n) < f_a(n_0).
\end{align}
Next, consider the case when the left-hand side of \eqref{fa_2} is nonnegative for some $n$. If $\a_{S^2}(n,r) > \a_{S^2}(n_0,r)$ for this $n$, and since by assumption $\a_{S^2}(n,r)$ grows without bounds as a function of $r$, we can always choose $r$ such that \eqref{fa_2} is violated, as the left-hand side is independent of $r$ while $\a_{S^2}(n,r) - \a_{S^2}(n_0,r)$ contains only terms proportional to positive powers of $r$. This forces us to conclude
\begin{align}
	\a_{S^2}(n,r) \leq \a_{S^2}(n_0,r), \quad\text{given} \quad f_a(n) \geq f_a(n_0).
\end{align}
Thus we see that $\a_{S^2}(n,r)$ has a local maximum at $n=n_0$. But $n_0$ is an arbitrary number, so this can be true only if $\a_{S^2}(n,r)$ is constant as a function of $n$. In this case, \eqref{fa_2} becomes
\begin{align}
	4f_a(n) - 4f_a(n_0) &\geq 0,
\end{align}
i.e. $f_a(n_0)$ is a local minimum. Again, since $n_0$ is arbitrary, this is possible only if
\begin{align}\label{fa_const}
	f_a(n) = \kappa
\end{align}
for some constant $\kappa$. However, we know from \cite{Lewkowycz:2014jia} that near $n=1$,
\begin{align}\label{fa_approx}
	f_a(n) \approx a - \frac{c}{2}(n-1) + \bigO\left((n-1)^2\right).
\end{align} 
Here, $a$ and $c$ are central charges defined via the conformal anomaly
\begin{align}
	\la T^\mu{}_\mu \ra = -\frac{a}{(4\pi)^2}E_4 + \frac{c}{(4\pi)^2}C_{\mu\nu\rho\sigma}C^{\mu\nu\rho\sigma},
\end{align}
and $E_4$ here is the Euler density. Thus, we see that near $n=1$, $f_a(n)$ certainly isn't constant if $c \not= 0$. This is indeed the case, as for 4d unitary CFTs we have \cite{Perlmutter:2015vma}
\begin{align}\label{ac_bound}
	\frac{1}{3} \leq \frac{a}{c} \leq \frac{31}{18}.
\end{align}
This contradicts \eqref{fa_const}, and thus we have successfully proved via contradiction that $\a_{S^2}(n,r)$ cannot grow without bounds as a function of $r$, assuming $S_n^\text{DR}$ admits a state-counting interpretation.

Indeed, one might have obtained the identical conclusion via the following intuitive argument.\footnote{We thank M. Headrick for bringing this argument to our attention.} The finite terms $\a_{\Sigma}(n,l_A)$ should in principle be removable by a counterterm, the relevant portion of which are integrals of local functions over the entangling surface $\Sigma$. On the other hand, they must be dimensionless, and the only finite length scale present is $l_A$. This suggests that $\a_{\Sigma}(n,l_A)$ may be topological and hence independent of the length scale $l_A$. Naturally this implies that in the case $\Sigma = S^2$, $\a_{S^2}(n,r)$ cannot grow without bounds as a function of $r$. 

\subsection{Constraint on $f_a(n)$}

We now return to examine inequality \eqref{fa_1}. As we are free to choose the radius of $\Sigma = S^2$, let us choose $r$ large enough so that
\begin{align}
	\a_{S^2}(n,r) \ll \log(\mu r).
\end{align}
This is possible using the conclusion obtained in the previous subsection above. In particular, in the limit when $r \to \infty$, we can write \eqref{fa_1} as
\begin{align}\label{final}
	f_a(n) \leq 0
\end{align}
for any $n > 1$. However, using \eqref{fa_approx} and the fact that both $a$ and $c$ are strictly positive \cite{Perlmutter:2015vma}, we see that \eqref{final} cannot be true for all $n$. This follows because using \eqref{ac_bound}, we can conclude that if $1 < n < \frac{5}{3}$, then
\begin{align}
	f_a(n) \approx  a - \frac{c}{2}(n-1) > 0,
\end{align} 
contradicting \eqref{final} and completing our proof that $S_n^\text{DR}$ does not admit a state-counting interpretation for 4d unitary CFTs.\footnote{It is interesting to note that the exact formula for $f_a(n)$ is determined for various theories and tabulated in \cite{Perlmutter:2015vma}. In each case, $f_a(n)$ is strictly positive, thus contradicting inequality \eqref{final} as required.}

\section{Conclusion}

We have demonstrated that a common method to regulate the infinities in the R\'enyi entropy of a 4d unitary CFT, namely dimensional regularization, gives us a renormalized R\'enyi entropy that lacks a state-counting interpretation. Therefore, another regularization scheme must be used if one wants to view the R\'enyi entropy of a 4d unitary CFT as the infinite-dimensional limit of a finite-dimensional quantum system. 

Other authors in the past have studied the connection between regulators and the state-counting interpretation of entanglement entropy \cite{Cooperman:2013iqr,Casini:2004bw,Jacobson:2012yt,Giaccari:2015vfh}. The arguments put forth by these authors suggest that the entanglement entropy for a QFT regulated by a covariant regularization scheme may not admit an underlying Hilbert space, and we provided a concrete proof that at least in the case of a 4d unitary CFT, there is no way to get around this if we choose to use dimensional regularization. 
We note here that while it is known in 2+1 dimensions that the renormalized entropy is negative, this does not a priori preclude the ability to prevent violation of positivity via the inclusion of counterterms. The analysis conducted in this note, however, allows one to eliminate that possibility, at least in the case of dimensional regularization. It would be of interest to study R\'enyi entropies of arbitrary QFTs obtained using more general covariant regulators, and see if one can obtain similar contradictions with the R\'enyi entropy inequalities.

\section*{Acknowledgements}

We would like to thank L. Bianchi, A. Lewkowycz, and S. Solodukhin for useful conversations. We would especially like to thank M. Headrick for reading the preprint and providing useful comments and feedback. N.B. is supported by the DuBridge Postdoctoral Fellowship, and also by the Institute for Quantum Information and Matter, an NSF Physics Frontiers Center (NFS Grant PHY-1125565) with support of the Gordon and Betty Moore Foundation (GBMF-12500028).

\bibliography{entropy-bib}{}
\bibliographystyle{unsrt}

\end{document}